\documentclass[sigconf]{acmart} 
\acmConference[SEAMS 2022]{The 17th Symposium on Software Engineering for Adaptive and Self-Managing Systems}{May 21–29, 2022}{Pittsburgh, PA, USA}

\usepackage[utf8]{inputenc}
\usepackage[inline]{enumitem}
\usepackage{diagbox}
\usepackage{graphicx}
\usepackage{pgfplots}
\usepackage{wrapfig}
\usepackage{amsmath}
\usepackage{caption}
\usepackage{subcaption}
\usepackage{hyperref}
\usepackage{algorithm}
\usepackage{algpseudocode}
\usepackage{verbatim}

\algnewcommand\algorithmicforeach{\textbf{for each}}
\algdef{S}[FOR]{ForEach}[1]{\algorithmicforeach\ #1\ \algorithmicdo}

\newcommand{\omgupdates}[1]{{\color{black} #1}}
\newcommand{\fordannycheck}[1]{{\color{black} #1}}

\title{Lifelong Self-Adaptation: Self-Adaptation Meets Lifelong Machine Learning}

\date{}

\author{Omid Gheibi}
\affiliation{%
    \institution{Katholieke Universiteit Leuven}
    \city{Leuven}
    \country{Belgium}
}
\email{omid.gheibi@kuleuven.be}

\author{Danny Weyns}
\affiliation{%
    \institution{Katholieke Universiteit Leuven, Belgium}
    \city{Linnaeus University}
    \country{Sweden}
}

\email{danny.weyns@kuleuven.be}

\acmYear{2022}
\acmConference[SEAMS '22]{17th International
Symposium on Software Engineering for Adaptive and Self-Managing Systems}{May
18--23, 2022}{PITTSBURGH, PA, USA}
\acmBooktitle{17th International Symposium on Software Engineering for Adaptive
and Self-Managing Systems (SEAMS '22), May 18--23, 2022, PITTSBURGH, PA, USA}
\acmDOI{10.1145/3524844.3528052}
\acmISBN{978-1-4503-9305-8/22/05}
\setcopyright{none}

\begin{document}

\begin{abstract}
In the past years, machine learning (ML) has become a popular approach to support self-adaptation. While ML techniques enable dealing with several problems in self-adaptation, such as scalable decision-making, they are also subject to inherent challenges. In this paper, we focus on one such challenge that is particularly important for self-adaptation: ML techniques are designed to deal with a set of predefined tasks associated with an operational domain; they have problems to deal with new emerging tasks, such as concept shift in input data that is used for learning. To tackle this challenge, we present \textit{lifelong self-adaptation}: a novel approach to self-adaptation that enhances self-adaptive systems that use ML techniques with a lifelong ML layer. The lifelong ML layer tracks the running system and its environment, associates this knowledge with the current tasks, identifies new tasks based on differentiations, and updates the learning models of the self-adaptive system accordingly. We present a reusable architecture for lifelong self-adaptation and apply it to the case of concept drift caused by unforeseen changes of the input data of a learning model that is used for decision-making in self-adaptation. We validate lifelong self-adaptation for two types of concept drift using two cases.  
\end{abstract}

\maketitle

\vspace{-5pt}
\section{Introduction}
\label{sec: introduction}

Self-adaptation equips a software system with a feedback loop that monitors the system and its environment and applies changes to the system when needed to realize a set of adaptation goals~\cite{Roadmap2009,weyns2020introduction}. Self-adaptation has been pivotal in automating tasks that otherwise need to be realized by operators~\cite{Kephart,garlan2004rainbow}, in particular tasks that are related to uncertainties that the system may face during its lifetime~\cite{esfahani2013usa,3487921}. 

Over the past years, we have observed an increasing trend in the use of machine learning (ML in short) to support self-adaptation. A recent systematic literature review~\cite{gheibi2021} shows that the number of articles published on this topic has been doubled every two years since 2014. For instance, ML has been used for efficient decision-making by reducing large adaption spaces~\cite{quin2019efficient}, for detecting abnormalities in the flow of activities in the  environment of the system~\cite{krupitzer2017adding}, and 
for learning changes of the system utility dynamically~\cite{8498142}.

While ML techniques enable dealing with several problems in software systems in general and self-adaptive systems in particular, these techniques are subject to several engineering challenges. Examples are the need for  specialized expertise when constructing ML solutions, providing reliable and efficient testing of ML techniques, dealing with unexpected events in the real operating environment of the system, and obtaining adequate quality assurances for ML applications~\cite{Kumeno2019,amershi2019software}. 

In this paper, we focus on one such challenge that is particularly important for self-adaptive systems: dealing with new learning tasks. In essence, ML techniques are designed to deal with a set of predefined tasks that they solve based on data derived from the operational domain of the system. Hence, they have problems to deal with new emerging tasks and changes in the operational domain. A typical example is new learning tasks that emerge due to unforeseen changes of input data over time. Without proper support, such problems require human intervention, which is time consuming and costly. This leads us to the research problem that we tackle in this paper:

\begin{quote}
    \textit{How to enable self-adaptive systems that use machine learning techniques to deal with new and changing learning tasks during operation?}
\end{quote}

\begin{sloppypar}
To tackle this research problem, we propose \emph{lifelong self-adaptation}: a novel approach to self-adaptation that enhances self-adaptive systems with a lifelong ML layer. The lifelong ML layer tracks the running system and its environment, associates the collected knowledge with the current tasks, identifies new tasks based on differentiations, and updates the learning models of the self-adaptive system accordingly. 
\end{sloppypar}

Lifelong self-adaptation leverages the principles of lifelong machine learning~\cite{thrun1998lifelong,chen2018lifelong} that offers an architectural approach for continual learning of a machine learning system.
Lifelong machine learning adds a layer on top of a machine learning system that 
selectively transfers the knowledge from previously learned tasks to facilitate the learning of new tasks within an existing or new domain~\cite{chen2018lifelong}. 
Lifelong machine learning has been successfully combined with a wide variety of learning techniques~\cite{chen2018lifelong}, including supervised~\cite{silver2015consolidation}, interactive~\cite{ammar2015autonomous}, and unsupervised learning~\cite{shu2016lifelong}.

Our focus in this paper is on self-adaptive systems that rely on architecture-based adaptation~\cite{KramerMagee,FORMS,Roadmap2009,Lemos2013}, where a self-adaptive system consists of a managed system that operates in the environment and a managing system that manages the managed system to deal with a set of adaptation goals. We focus on managing systems that comply with the MAPE-K reference model, short for  Monitor-Analyse-Plan-Execute-Knowledge~\cite{Kephart,empiricalMAPE}. Our particular focus is on  managing systems that use a ML technique to support any of the MAPE-K functions. We make the assumption that dealing with new learning tasks does not require any runtime evolution of the software of the managed and managing system; hence we focus at handling new learning tasks that is realized by evolving the learning models used by the managing system. 

We study one concrete instance of new learning tasks in \omgupdates{self-adaptation}: concept drift. Concept drift~\cite{webb2016characterizing} refers to an unforeseen change of the input data of a learning model that results in predictions becoming less accurate over time, which may jeopardize the reliability of the system. We look at two types of concept drift: sudden and incremental concept drift.  

The concrete contributions of this paper are: 
\begin{enumerate}
    \item A reusable architecture for lifelong self-adaptation; 
    \item Two concrete instances of the architecture to deal with sudden and incremental concept drift respectively;
    \item A validation of the two instances of the architecture using DeltaIoT~\cite{iftikhar2017deltaiot} and a gas delivery system~\cite{VERGARA2012320} respectively.
\end{enumerate}

The remainder of this paper is structured as follows. In Section~\ref{sec:preliminaries}, we provide a brief introduction of lifelong machine learning that provides the basic framework underlying lifelong self-adaptation.
Section~\ref{sec:lifelong-self-adaptation} then introduces the novel approach for lifelong self-adaptation. 
Next, we instantiate the architecture of a lifelong self-adaptive system for sudden concept drift in Section~\ref{sec:lifelong-self-adaptation-sudden-concept-drift} using DeltaIoT for the evaluation. Section~\ref{sec:lifelong-self-adaptation-incremental-concept-drift} instantiates the architecture for incremental concept drift using a gas delivery case. In Section~\ref{sec:validity} we discuss threats to validity, and  
Section~\ref{sec:related-work} presents related work. Finally, we wrap up and outline opportunities for future research in  Section~\ref{sec:conclusions}. 

\vspace{-4pt}
\section{Lifelong Machine Learning in a Nutshell}\label{sec:preliminaries}

We briefly describe the basic principles of lifelong machine learning that provide the basis for  lifelong self-adaptation. 

Lifelong machine learning is the ability of a machine-learning system to learn new tasks that were not predefined when the system was designed~\cite{THRUN199525}.
Technically, lifelong machine learning is a continuous learning process of a learner~\cite{chen2018lifelong}. Assume that at some point in time, the learner has performed a sequence of $n$ learning tasks, $\mathcal{T}_1, \mathcal{T}_2,\cdots, \mathcal{T}_n$, called the \textit{previous tasks}, that have their corresponding data sets $\mathcal{D}_1 , \mathcal{D}_2 ,\cdots, \mathcal{D}_n$. Tasks can be of different \textit{types} and from different \textit{domains}. When faced with 
task $\mathcal{T}_{n+1}$ (called the \textit{new} or \textit{current task}) with its data set  $\mathcal{D}_{n+1}$, the learner can leverage \textit{past knowledge} maintained in a \textit{knowledge-base} to help learn task  $\mathcal{T}_{n+1}$. The new task may be given by a stakeholder or it may be discovered automatically by the system. 
Lifelong machine learning aims to optimize the performance of the learner for the new task (or for an existing task by treating the rest of the tasks as previous tasks). After completing learning task $\mathcal{T}_{n+1}$, the knowledge base is updated with the new gained knowledge, e.g., using intermediate and ﬁnal results obtained via learning. Updating the knowledge can involve checking consistency, reasoning, and mining meta-knowledge. 
    
For example, consider a lifelong machine learning system for the never-ending language learner~\cite{mitchell2018never} (NELL in short). NELL aims at answering questions posed by users in natural language. To that end,  it sifts the Web 24/7 extracting facts, e.g., ``Paris is a city." 
The system is equipped with a set of classifiers and deep learners to categorize nouns and phrases, e.g., ``apple'' can be classified as ``Food'' and ``Company'' falls under an ontology, and detecting relations, e.g., ``served-with'' in ``tea is served with biscuits.''
NELL can infer new beliefs from this extracted knowledge, and based on the recently collected web documents, NELL can expand relations between existing noun phrases or the ontology. This expansion can be a change within existing ontological  domains, e.g., politics or sociology, or be a new domain like internet-of-things. Hence, the expansion causes an emerging task like classifying new noun phrases for the expanded part of the ontology. 

Lifelong machine learning works together with different types of learners. In lifelong supervised learning, every learning task aims at recognizing a particular class or concept. E.g., in cumulative learning, identifying a new class or concept is used to build a new multi-class classifier for all the existing and new classes using the old classifier~\cite{2939672.2939835}. Lifelong unsupervised learning focuses on topic modeling and lifelong information extraction, e.g., by mining knowledge from topics resulting from previous tasks to help generating better topics for new tasks~\cite{2872427.2883086}. In lifelong semi-supervised learning, the learner enhances the number of relationships in its knowledge base by learning new facts, for instance, in the NELL system~\cite{mitchell2018never}. Finally, in lifelong reinforcement learning each environment is treated as a task~\cite{Tanaka1998AnAT},
or a continual-learning agent solves  complex tasks by learning easy tasks first~\cite{Ring1997}. Recently, lifelong learning has gained increasing attention, in particular for autonomous learning agents and robots based on neural networks~\cite{abs-1802-07569}. 

One of challenges for lifelong machine learning is dealing with catastrophic forgetting, i.e., the loss of what was previously learned while learning new information, which can lead to system failures~\cite{abs-1908-01091}.  Another challenge for any machine learning pipeline is under-specification, i.e., a significant decrease of the  performance of a learning model from training to  deployment (or testing)~\cite{d2020underspecification}.
Promising approaches have been proposed, see e.g.,~\cite{abs-1802-07569} for catastrofic forgetting or~\cite{ribeiro-etal-2020-beyond} for under-specification. Yet, more research is required to transfer these techniques to real-world systems.

\begin{figure*}[t!]
    \centering
     \includegraphics[width=0.93\linewidth]{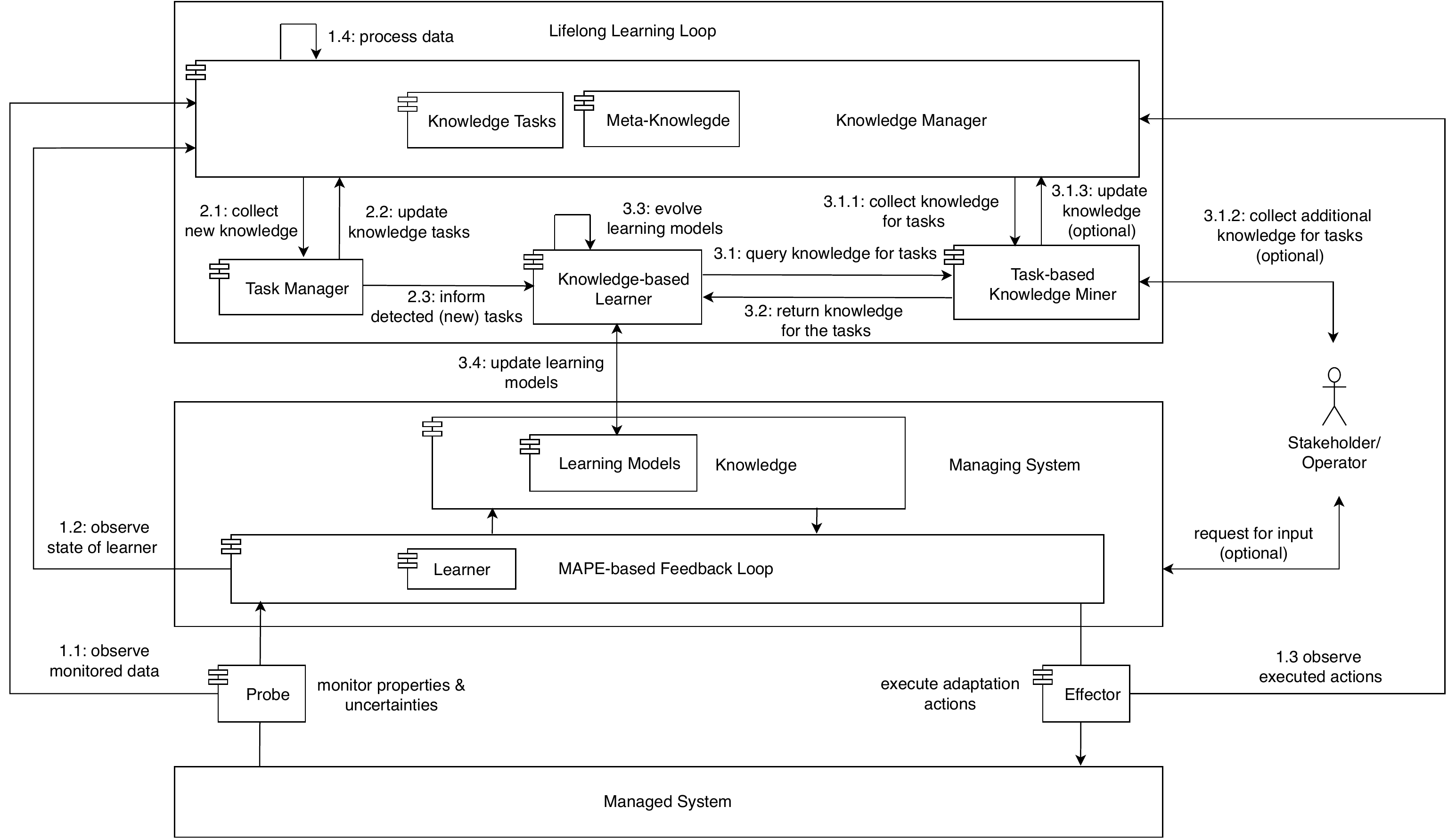}
    \caption{The architecture of a lifelong self-adaptive system
    \label{fig:lsa}\vspace{-5pt}}
\end{figure*}

\vspace{-5pt}
\section{Lifelong Self-Adaptation}
\label{sec:lifelong-self-adaptation}
We now introduce the novel approach of lifelong self-adaptation that enables  self-adaptive systems that use machine  learning techniques to deal with new learning tasks during operation. We start with assumptions and requirements for lifelong self-adaptation. Then we present the architecture of a lifelong self-adaptive system, explain how it deals with new tasks, and introduce instances for two types of concept drift. 

\subsection{Assumptions for Lifelong Self-Adaptation}
 
\begin{sloppypar}
The assumptions that underlie the approach for lifelong self-adaptation presented in this paper are: 
\end{sloppypar}

\begin{itemize}
    \item The self-adaptive system comprises a managed system that \omgupdates{realizes} the domain goals for users in the environment, and a managing system that interacts with the managed system and realizes the adaptation goals; 
    \item The managing system is equipped with a learner that supports the realization of self-adaptation; 
    \item \fordannycheck{The self-adaptive system provides the probes to collect the data that is required for realizing lifelong self-adaptation}; this may include an interface to the user to support the lifelong self-adaptation process if needed;
    \item The managing system provides the necessary interface to adapt the learning models.  
\end{itemize}

In addition, we only consider new learning tasks that require evolution of the learning models; runtime evolution of the software of the managed or managing system is out of scope.

\subsection{Requirements for Lifelong Self-Adaptation}

The lifelong self-adaptive systems should:  

\begin{enumerate}
    \item[R1] Provide the means to collect and manage the data that is required to deal with new tasks; 
    \item[R2] Be able to discover new tasks based on the collected data; 
    \item[R3] Be able to determine the required evolution of the learning models to deal with the new tasks; 
    \item[R4] Evolve the learning models such that they can deal with the new tasks.   
\end{enumerate}

\vspace{-5pt}
\subsection{Architecture of Lifelong Self-Adaptive Systems}

Figure~\ref{fig:lsa} shows the architecture of a lifelong self-adaptive system. We zoom in on the role of each component and explain the flow of activities to deal with new and changing tasks.

\paragraph{Managed System} Takes input from the environment and realizes the domain goals for the users of the system. 

\paragraph{Managing System} Monitors the managed system and environment and executes adaptation actions on the managed system to realize the adaptation goals. The managing system comprises the MAPE components that share knowledge.
A learner supports the MAPE functions. The learning models are stored in the knowledge. The system may request an operator for input (active learning). 

\paragraph{Lifelong Learning Loop} Adds a meta-layer on top of the managing system, leveraging the principles of lifelong machine learning. This layer tracks the layers beneath and when it detects a new learning task, it will evolve the learning models of the learner accordingly. We elaborate now on the components of the lifelong learning loop and their interactions.

\paragraph{Knowledge Manager} Stores all knowledge that is relevant to the learning tasks of the learner of the managing system (realizing requirement R1). In each adaptation cycle, the knowledge manager collects a knowledge triplet: $k_i$\,=\,$\langle$input$_i$,\,state$_i$,\,output$_i$$\rangle$. Input are the  properties and uncertainties of the system and its environment (activity 1.1). State refers to data of the managing system relevant to the learning tasks, e.g., settings of the learner (1.2). Output refers to the actions applied by the managing system to the managed system (1.3). Sets of knowledge triplets are labeled with tasks, i.e., $\langle$t$_i$, $\{$k$_u$,\,k$_v$,\,k$_w$$\}$$\rangle$, a responsibility of the task manager. The labeled triplets are stored in the knowledge tasks repository.  

Depending on the domain, the knowledge manager may reason about new knowledge or mine the knowledge  extracting (or updating) meta-knowledge, such as a cache or an ontology (1.4). The meta-knowledge can be used by the other components of the lifelong learning loop to enhance their performance. The knowledge manager may synthesize parts of the knowledge to manage the amount of stored knowledge (e.g., outdated or redundant tuples may be marked or removed).

\paragraph{Task manager} Is responsible for detecting new learning tasks (realizing R2). The task manager periodically retrieves new knowledge triplets from the knowledge manager (2.1). The duration of a period is domain-specific and can be one \omgupdates{or} more adaptation cycles of the managing system. The task manager then identifies task labels for the retrieved knowledge triplets. A triplet can be assigned the label of an existing task or a new task. Each new task label represents a (statistically) significant change in the data of the knowledge triplets, e.g., a significant change in the distribution of the data observed from the environment and managed system.
Hence, a knowledge triplet can be associated with multiple task labels, depending on the overlap of their corresponding data (distributions).  
The task manager then returns the knowledge triplets with the assigned task labels to the knowledge manager that updates the knowledge accordingly (2.2). Finally, the task manager informs the knowledge-based learner about the new tasks (2.3).

\paragraph{Knowledge-based learner} Decides how to evolve the learning models of the learner of the managing system based on the collected knowledge and associated tasks (realizing R3), and then enacts the evolution of the learning models (realizing R4). To collect the knowledge it needs for the detected learning tasks, the knowledge-based learner queries the task-based knowledge miner (3.1) that returns task-specific data (3.2); the working of the task-based knowledge \omgupdates{miner} is explained below.
The knowledge-based learner then uses the collected data to evolve the learning models of the managing system (3.3). This evolution is domain-specific and depends on the type of learner at hand, e.g., tuning or retraining the learning models for existing tasks, or generating and training new learning models for newly detected tasks. Finally, the knowledge-based learner updates the learning models (3.4).

\paragraph{Task-based knowledge miner} Is responsible for collecting the data that is required for evolving the learning models for given learning task by the knowledge-based learner (supports realizing R3). As a basis, the task-based knowledge miner retrieves the knowledge triplets associated with the given task from the knowledge tasks repository, possibly exploiting meta-knowledge, such as a cache (3.1.1). Additionally, the task-based knowledge miner can mine the knowledge repositories of the knowledge manager, e.g., to retrieve knowledge of learning tasks that are related to the task requested by the knowledge-based learner. Optionally, the task-based knowledge miner may collect knowledge from stakeholders, for instance to confirm labels of knowledge triplets (3.1.2). Finally, the miner uses new knowledge to update the knowledge maintained in the repositories by the knowledge manager (3.1.3), e.g., it may update meta-knowledge about related tasks or add data to the knowledge provided by stakeholders. For the use cases in the next sections, we do not consider advanced reasoning and mining by the task-based knowledge \omgupdates{miner}. 

\vspace{-2pt}
\subsection{Lifelong Self-Adaptation for Concept Drift}

The concrete problem of new learning tasks we tackle with lifelong self-adaptation in this paper is concept drift. Concept drift refers to a change of the statistical characteristics of data over time~\cite{basseville1993detection,webb2016characterizing}, which may deteriorate the performance of the learning models that use that data to learn. Our focus is on co-variate drift, a common form of concept drift, that arises when the distributions of labeled data attributes change over time and the distribution of target values remains invariant relative to the attributes~\cite{gama2014survey,webb2016characterizing}. We study two types of drift: sudden co-variate drift and incremental co-variate drift. The next sections instantiate the architecture for lifelong self-adaptation for these types and apply each solution to a case.

\section{Lifelong Self-Adaptation to Deal with Sudden Co-variate Drift}
\label{sec:lifelong-self-adaptation-sudden-concept-drift} 
\begin{figure}[!h]
    \centering
    \includegraphics[width=\linewidth]{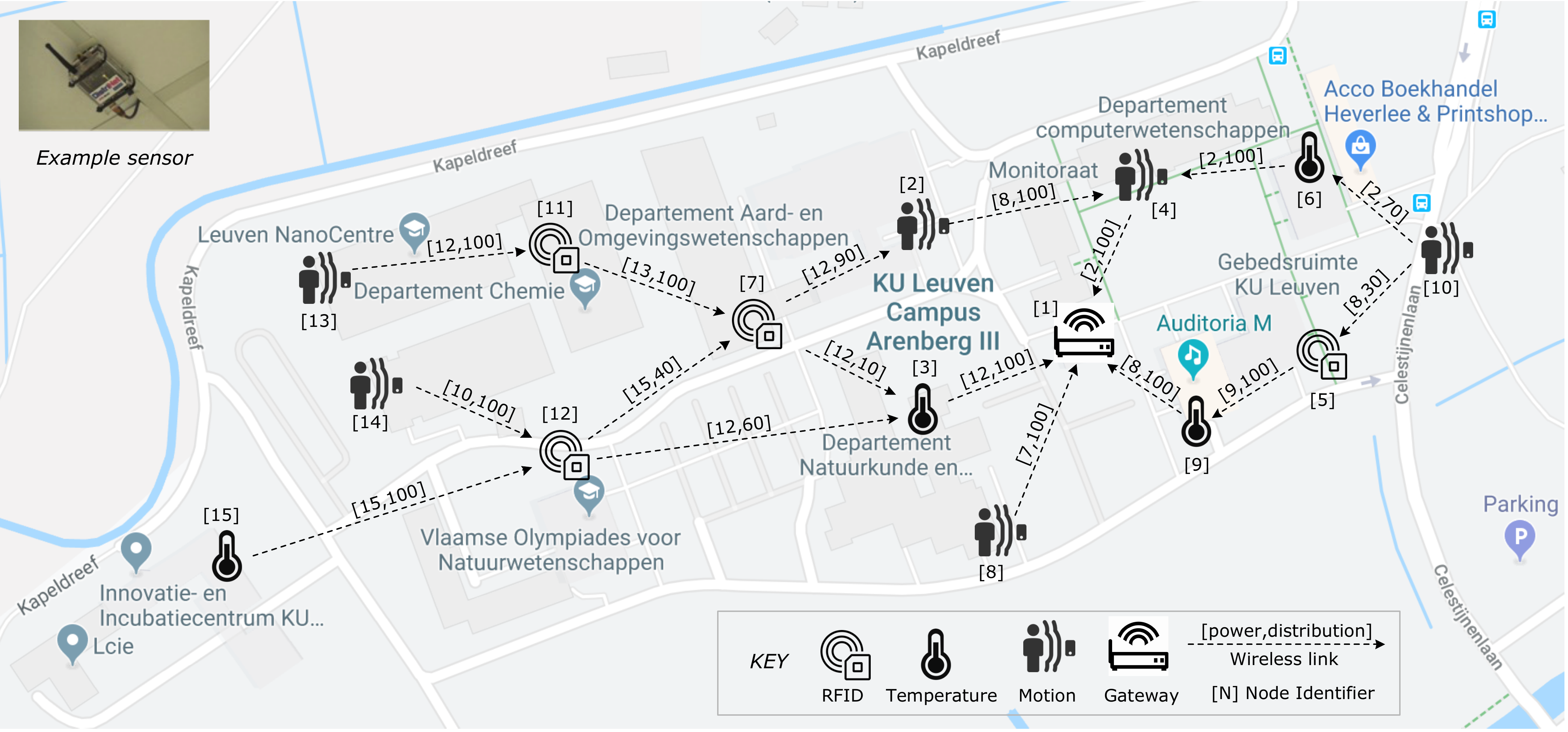}
    \caption{DeltaIoT setup used for the evaluation~\cite{quin2019efficient}
    \label{fig:deltaiot-setup}}
\end{figure}

To explain how lifelong self-adaption deals  with sudden co-variate drift we use the DeltaIoT artifact~\cite{iftikhar2017deltaiot}. Figure~\ref{fig:deltaiot-setup} shows the setup of the IoT network that consists of 15 battery-powered sensor motes that measure parameters in the environment and send the data via a wireless multi-hop network to a central gateway where users can use the data. The communication in the network is time-synchronized~\cite{mills2017computer}, i.e., neighboring motes are allocated slots where they can send and receive messages over a wireless link as shown in the figure. 
\vspace{-5pt}

\begin{figure}[!h]
    \centering
    \includegraphics[scale=0.38]{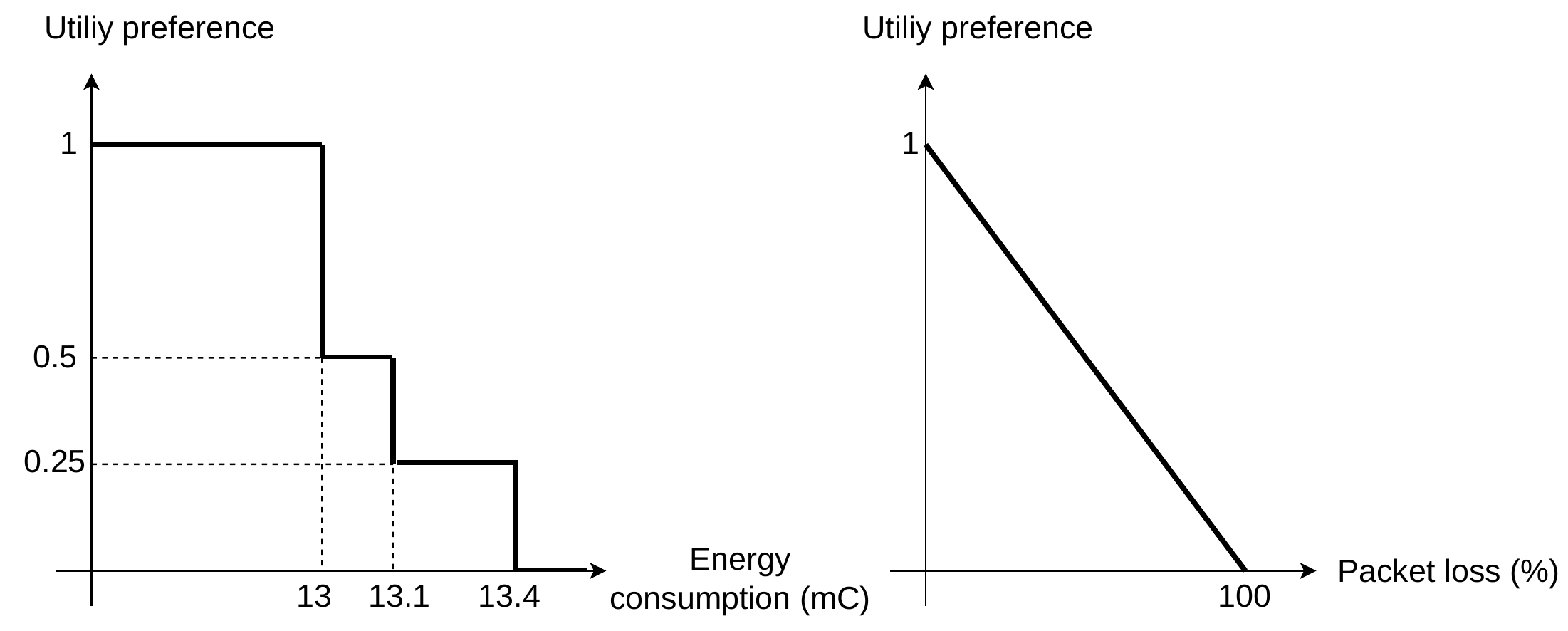}
    \caption{Utility preferences for qualities in DeltaIoT 
    \label{fig: utility preferences}}
\end{figure}

The quality properties of interest in this paper are packet loss and energy consumption. Figure~\ref{fig: utility preferences} shows the utility preferences of the stakeholders for the qualities of interest. 

The utility $U_c$ of a network configuration is defined as:

\begin{equation}
    U_c = 0.2 \cdot p_{ec} + 0.8 \cdot p_{pl}
\end{equation}

with $p_{ec}$ and $p_{pl}$ the utility for energy consumption and packet loss respectively, and $0.2$ and $0.8$ the weights associated with the quality properties. Users give max preference to an energy consumption below 13~mC (milli Coulomb) and zero preference to energy consumption above 13.4~mC. The utility for packet loss decreases linearly between 0\% and 100\%. 

The quality properties are affected by uncertainties. We consider two types: \emph{network interference} caused by environmental conditions such as weather fluctuations, and \emph{load of messages} generated by motes that fluctuate based on different factors such as the presence of people in the environment. The network interference affects the Signal-to-Noise Ratio (SNR)~\cite{haenggi2009stochastic} that determines packet loss. 

To maximize the utility, we add a managing system at the gateway that monitors the managed system and its environment and adapts the networks settings in each cycle. The \emph{power setting} of each node can be set (0 and 15), which will affect the SNR and hence the packet loss, and 
the \emph{distribution of the messages} along the network links can be set (for motes with two links there settings are: $0/100$, $20/80$, $40/60$, $60/40$, $80/20$, $100/0$). In total, there are 216 possible configurations. 

The managing system decides about whether or not to adapt the system after each communication cycle (8 minutes in our scenario). To select a configuration, the managing system uses a learner that predicts the quality attributes of the adaptation options (i.e., the possible configurations), leveraging  on~\cite{quin2019efficient}. Concretely, we use two settings for the evaluation: a basic setting with a stochastic gradient descent (SGD) regressor~\cite{saad1998online}, and an additional setting that allows switching between a SGD and a Hoeffding Adaptive Tree (HAT)~\cite{HAT}. An SGD regressor estimates the gradient of the loss for each data sample and uses that to update the learning model with a decreasing learning rate. A HAT regressor is an adaptive Hoeffding decision tree that incrementally grows in different branches by detecting different distributions in data.
The regressors use feature vectors that comprise the data of an adaptation option (power settings and links' distributions) and the monitored uncertainties (SNR's of links and loads of messages). Based on these predictions, the configuration with the highest utility is selected and used to adapt the network configuration. 

\subsection{Problem}

\begin{figure}[!h]
    \centering
    \includegraphics[scale = 0.4]{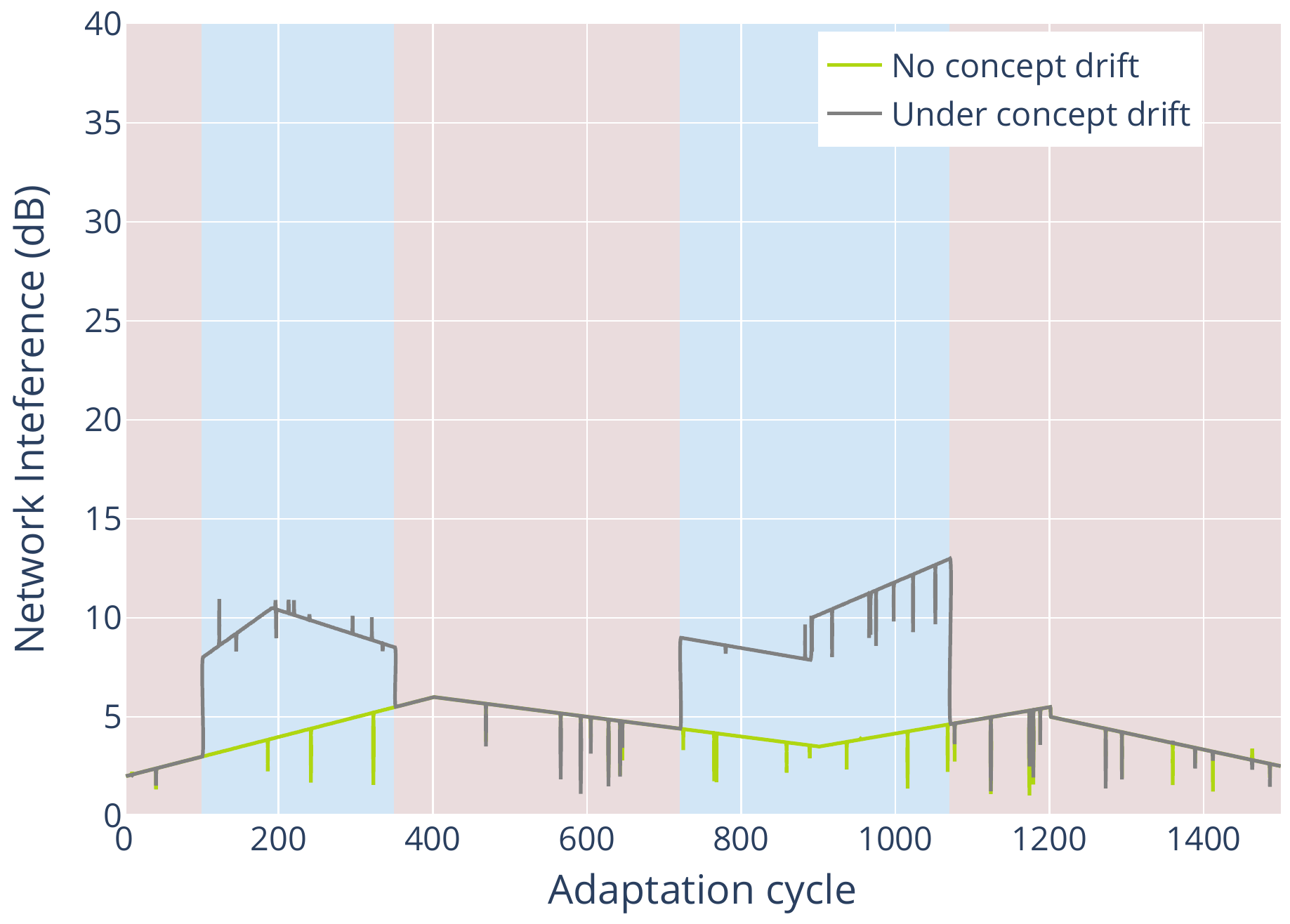}
    \caption{Global network interference in DeltaIoT; no concept drift (green) vs. concept drift occurs recursively (gray)}
    \label{fig: network interference drift}
\end{figure}

The problem is now that the distributions of input data that is used by the regressors may drift over time. Figure~\ref{fig: network interference drift} shows how the global network interference changes when sudden concept drift occurs. Such a concept drift may for instance occur due to construction works in the neighbourhood of the IoT network where some machinery may periodically produce extra noise affecting the SNR of the network links~\cite{din2017experimental}.

\begin{figure}[!h]
    \centering
    \includegraphics[scale=0.4]{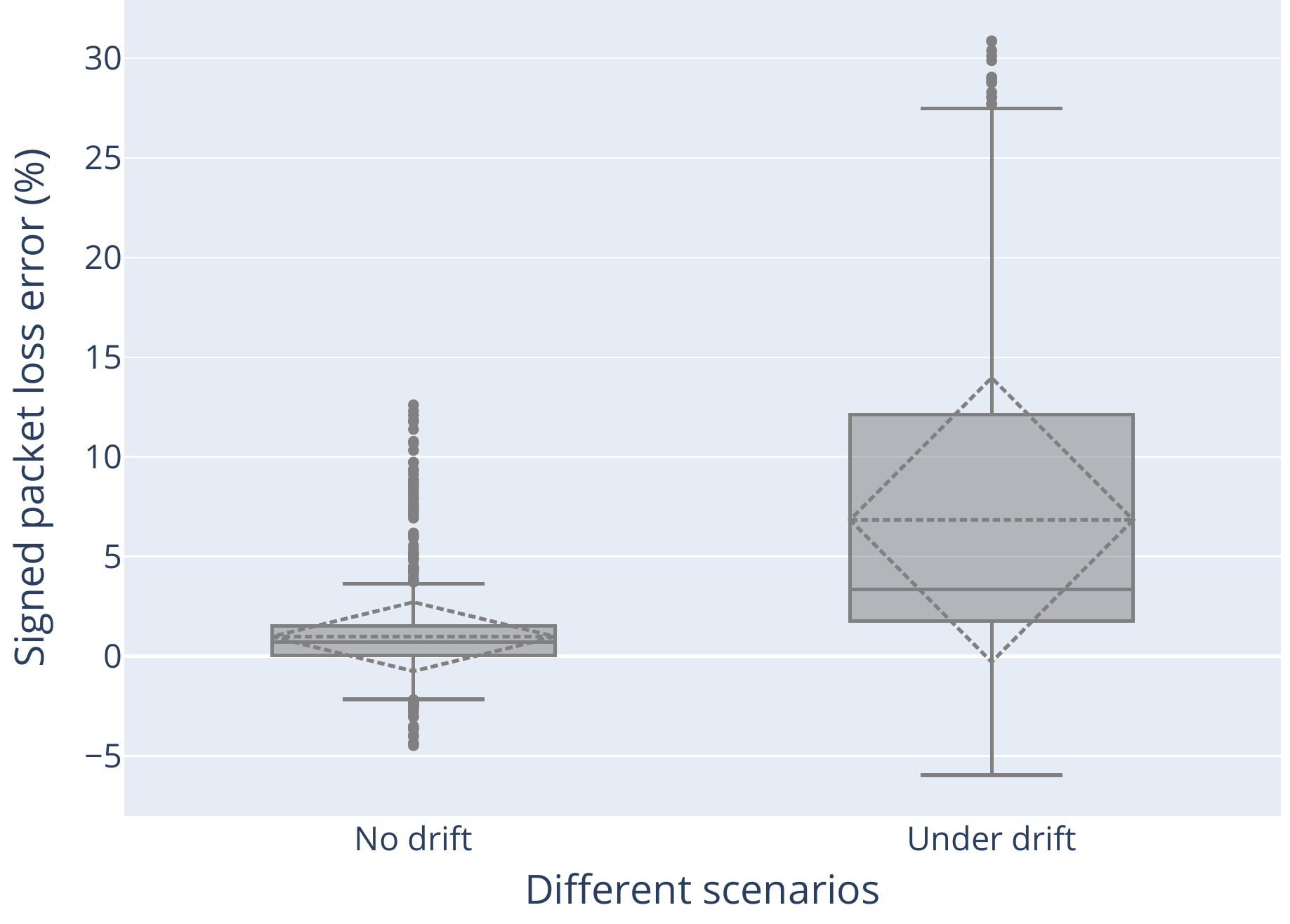}
    \caption{Distributions of the signed differences of packet loss between the true best options and the adaptation options selected by a SGD regressor with and without drift. 
    \label{fig: pl distribution drift vs no drift}}
\end{figure}

Figure~\ref{fig: pl distribution drift vs no drift} shows the distribution of the signed difference between the packet loss predicted by a SGD regressor (without lifelong self-adaptation) and the true best options that are based on the true values of the quality properties. We used here a scenario of 1500 adaptation cycles without and with drift.

The median of the signed error increases from 0.70\% to 3.36\% and the mean from 0.98\% to 6.84\%, pointing to a significant increase in packet loss.  
The underlying problem is that concept drift introduces new emerging learning tasks that the regressors cannot handle. 
To tackle this problem, we add a lifelong learning loop on top of the managing system (instantiating the lifelong learning loop shown in Figure~\ref{fig:lsa}).

\subsection{Lifelong Learning Loop Instance}
\label{sec:Lifelong learning loop instance}
The lifelong learning loop
is triggered every 20 adaptation cycles of the managing system (empirically determined). 

\subsubsection{Knowledge Manager} Collects 20 new knowledge triplets per cycle of the lifelong learning loop. Input of the triplets are packet loss and energy consumption of the system (i.e., system properties), and SNR's for all links and the load of messages for each mote (i.e., uncertainties). 
State is meta-data about learner, including the configuration of the learner such as the scaler used, the input vectors that include the adaptation options, and the corresponding target values predicted by the learning model. 
Output are the adaptation actions applied for the selected adaptation option. 
In this instance, the knowledge manager does not maintain any meta-knowledge. 

\subsubsection{Task Manager}
Detects new tasks (i.e., new distribution of features in the input data of knowledge triplets). We use  auto-encoders~\cite{jaworski2020concept,yang2021cade, andresini2021insomnia}. An auto-encoder is an artificial neural network that learns to encode unlabeled data. 
Suppose at some point in time, the system has knowledge about a set of previously detected tasks denoted by $\mathcal{T}_1, \ldots, \mathcal{T}_n$. 
These tasks have  corresponding data sets $\mathcal{D}_1, \ldots, \mathcal{D}_n$, each data set consisting of knowledge triplets\footnote{To speed up detection, we use 100 triplets; empirically determined.}, and their corresponding auto-encoders $\mathcal{AE}_1, \ldots, \mathcal{AE}_n$.
When triggered, the task manager determines the statistical similarity\footnote{We use the p-value from a non-parametric  Mann-Whitney-U test~\cite{mann1947test}.} between the outputs of each $\mathcal{AE}_i$ on the inputs of the newly observed knowledge triplets and the outputs of $\mathcal{AE}_i$ on  $\mathcal{D}_i$.
If this similarity is below a given threshold\footnote{We use p=0.025 to ensure high confidence of the decision. \omgupdates{Note that if we denote p-values in ascending order $p_1, p_2, \ldots, p_n$, the threshold $p$ will be adjusted by Holm's correction method~\cite{holm1979simple} to $\frac{p}{n-i+1}$ for $p_i$.}} for all auto-encoders, the task manager introduces a new task $\mathcal{T}_{n+1}$ with the new triplets as $\mathcal{D}_{n+1}$.
Then the task manager instantiates the auto-encoder $\mathcal{AE}_{n+1}$ (i.e., selecting the hyper-parameters: number of layers, neurons, etc.) for which we use a Bayesian optimizer~\cite{omalley2019kerastuner}.  
If the similarity is greater than the threshold for at least one auto-encoder, the task manager assigns label  $\mathcal{T}_i$ to corresponding triplets based on the highest achieved similarity of $\mathcal{AE}_{i}$. In the evaluation settings, we consider only the case where a task manager detects one new task per cycle.

\subsubsection{Knowledge-Based Learner} 
Evolves the learning models. 
The function the knowledge-based learner is shown in Algorithm~\ref{algo:knowledge based learner}. When informed about a detected (new) task (Line~\ref{line: detected task id}), the knowledge-based learner queries appropriate knowledge specific to the task from the task-based knowledge miner (Line~\ref{line: collect approp knowledge from TBKM}, see the next part). Based on the collected data, the knowledge-based learner determines the best learning models (Line~\ref{line: best models start} until Line~\ref{line: best models end}). 
The knowledge-based learner optimizes the hyper-parameters of each learning model using a Bayesian optimizer (Line~\ref{line: optimising model}).
The knowledge-based learner then trains the best model based on the collected training data (Line~\ref{line: train the best model}). When all learning models are trained, they are updated in the knowledge repository of the feedback loop (Line~\ref{line: update learning models}). 

\subsubsection{Task-Based Knowledge Miner} Fetches the sets of knowledge triplets from the knowledge manager for the tasks  queried by the knowledge-based learner. We limit the fetched data\footnote{We use the last 1000 task-specific data; empirically determined.}  to keep the training time of the knowledge-based learner at a constant and appropriate level. The current instance of the task-based knowledge miner applies no special mining. 

\begin{algorithm}[!t]
    \caption{Function Knowledge-Based Learner (per task)}
    \label{algo:knowledge based learner}
    \begin{algorithmic}[1]
        \State $\textit{detected\_task\_id} \gets $ \textit{From Task Manager} 
        \label{line: detected task id}
        \State $\textit{queried\_knowledge}$ $\gets$ $\textit{Task\_Based\_Knowledge\_Manager.}
        \newline
        \hspace*{3em}\text{query\_knowledge\_for\_task}\textit{(}$$\textit{detected\_task\_id)}$
        \label{line: collect approp knowledge from TBKM}
        \State $\textit{best\_models}\gets\textit{[]}$
        \label{line: best models start}
        \ForEach{$\textit{learning\_model} \in
        \newline \hspace*{3em}\textit{queried\_knowledge[state.learning\_models]}$}
            \State 
            $\textit{training\_data}\gets\textit{ queried\_knowledge[input,output]}$ 
            \label{line: get model target}
            \State 
            $\textit{hypparam}\gets\textit{ learning\_model.hyper\_param}$ 
            \label{line: get model target}
            \label{line: hyper-parameter tuning for SGD and HAT}
            \State $\textit{best\_model}\gets\text{bayesian\_optimizer}\textit{(hypparam,}$
            \newline \hspace*{15em} $\textit{training\_data)}$ 
            \label{line: optimizing model}
            \State $\textit{best\_model}\text{.train}\textit{(training\_data)}$
            \label{line: train the best model}
            \State $\textit{best\_models}\text{.add}\textit{(best\_model)}$
        \EndFor  \label{line: best models end}
    \State $\textit{Feedback\_Loop.Knowledge}$
    \newline \hspace*{3em}$\text{.update\_learning\_models}\textit{(best\_models)}$   \label{line: update learning models}  
    \end{algorithmic}
\end{algorithm}

\subsection{Evaluation Results}
We validated the instance of the lifelong self-adaptation architecture using the DeltaIoT simulator. We used the setup presented above (see Figure~\ref{fig:deltaiot-setup}) and applied a scenario over 1500 adaptation cycles, representing 10 days wall clock time,  with and without concept drift (see Figure~\ref{fig: network interference drift}). The evaluation was done on a computer with an i7-3770 @ 3.40GHz processor and 16GB RAM. All material of the evaluation is available at the project website~\cite{project-website}. 

We compared five approaches: (1) an optimal approach that selects the best adaptation option by maximizing the expected utility $U_c$ based on the true values of the quality attributes in each adaptation cycle (referred to as Baseline), (2) a state-of-the-art approach~\cite{chen2019all} that uses a regressor that retrains learning models (SGDs) in each adaptation cycle with all previously collected data (State-of-the-art), (3) incremental learning with a SGD regressor but without a lifelong learning loop (SGD without LLL), (4) lifelong self-adaptation with a lifelong learning loop and a SGD regressor (SGD with LLL), and finally (5) lifelong self-adaptation with a lifelong learning loop that can switch the learner of the managing system between a SGD regressor and HAT (SGD + HAT with LLL). This last approach allows comparing the effect of incorporating HAT, a learning method that is known to be suitable for dealing with concept drift of input data. 

\begin{figure}[!t]
    \centering
    \includegraphics[scale=0.4]{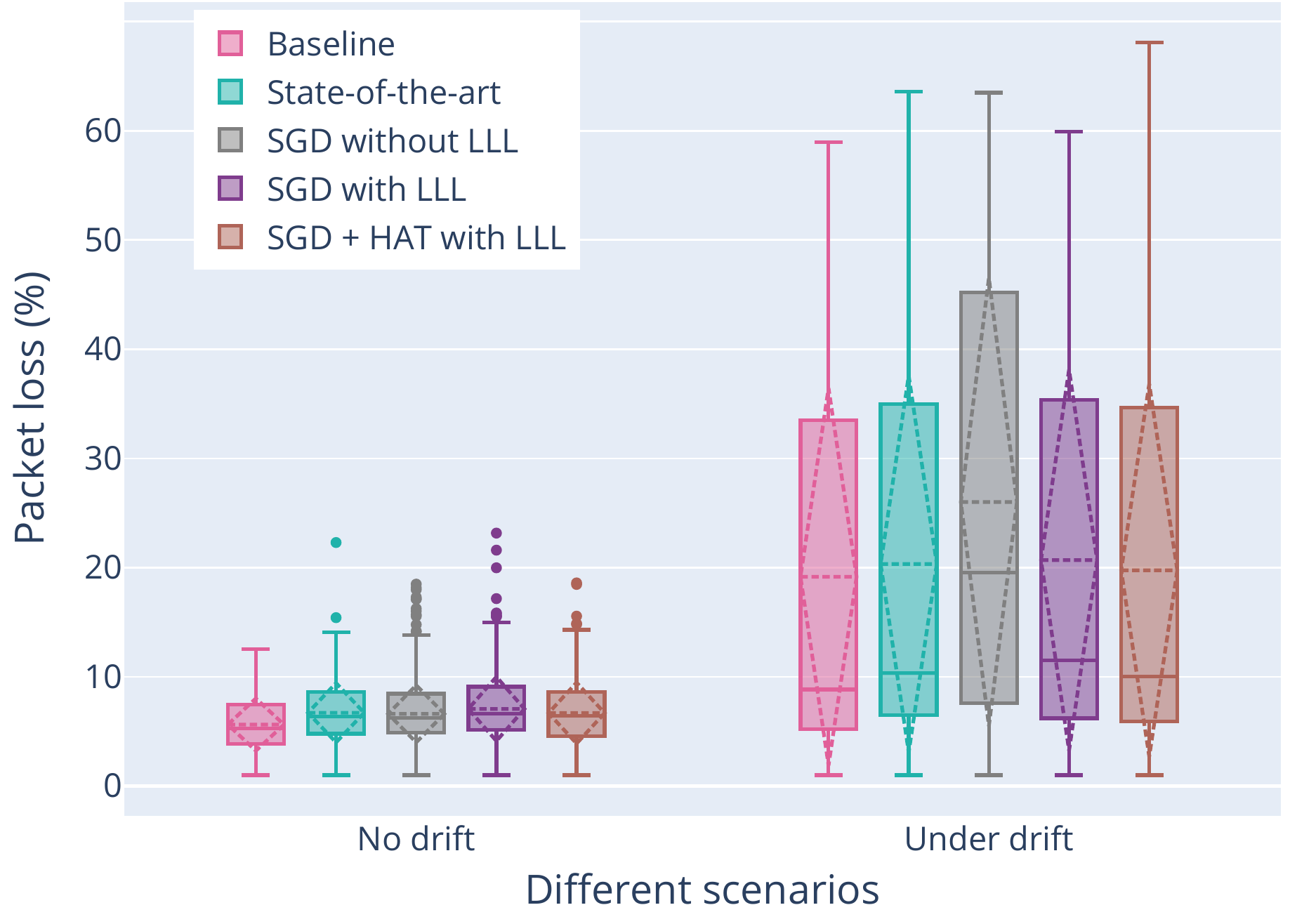}
\caption{Packet loss for different methods with/without drift 
    \label{fig:packet-loss-deltaiot}\vspace{-15pt}}
\end{figure}

We investigated the following two evaluation questions: 

\begin{enumerate}
    \item To what extent can lifelong self-adaptation deal with concept drift of input data for its learner on the  quality properties and system utility compared to the baseline? 
    \item How does lifelong self-adaptation compare with the state-of-the-art approach presented in~\cite{chen2018lifelong}?  
\end{enumerate}

\subsubsection{Impact on quality properties and system utility}
Figure~\ref{fig:packet-loss-deltaiot} summarizes the results for packet loss of the network with the different approaches. 
The results show that all approaches perform equally well for a scenario without concept drift (boxplots on the left hand side). For the scenario with concept drift (boxplots on the right hand side), the median of packet loss for the baseline approach (optimal approach based on the true values of quality properties) increases from 5.25\% to 8.84\%, indicating an inherent impact of the additional network interference. However, for a managing system that uses a SGD regressor (without a lifelong learning loop) the values increase from 6.23\% to 19.56\%. These results confirm earlier findings of Chen~\cite{chen2019all} when applying incremental learning methods under concept drift in different domains, such as service-based systems and cloud.
The dramatic increase of packet loss (combined with the effects on energy consumption) results in a substantial decrease of 12\% in the median of the expected utility of the IoT network (for detailed results of energy consumption and utility, see the project website~\cite{project-website}). 

In contrast, when applying lifelong self-adaptation (LLL), the results are similar to the baseline: the median of packet loss is slightly higher, 2.69\% for SGD with LLL and 1.18\% for SGD + HAT with LLL (compared to 10.72\% for SGD without a lifelong learning loop). The expected utility for the lifelong learning approaches is only marginally smaller compared to the baseline: 1\% and 2\% for SGD with LLL and SGD + HAT with LLL respectively (compared to 12\% for SGD without LLL). We observe that the performance of the two lifelong self-adaptation approaches are very similar, so the effect of including HAT as alternative for SGD is negligible.  

In answer to the first evaluation question, we conclude that lifelong self-adaptation can effectively deal with sudden concept drift of input data for its learner on the  qualities and system utility and performs similar to the baseline.  

\begin{figure}[!t]
    \centering
    \includegraphics[scale=0.4]{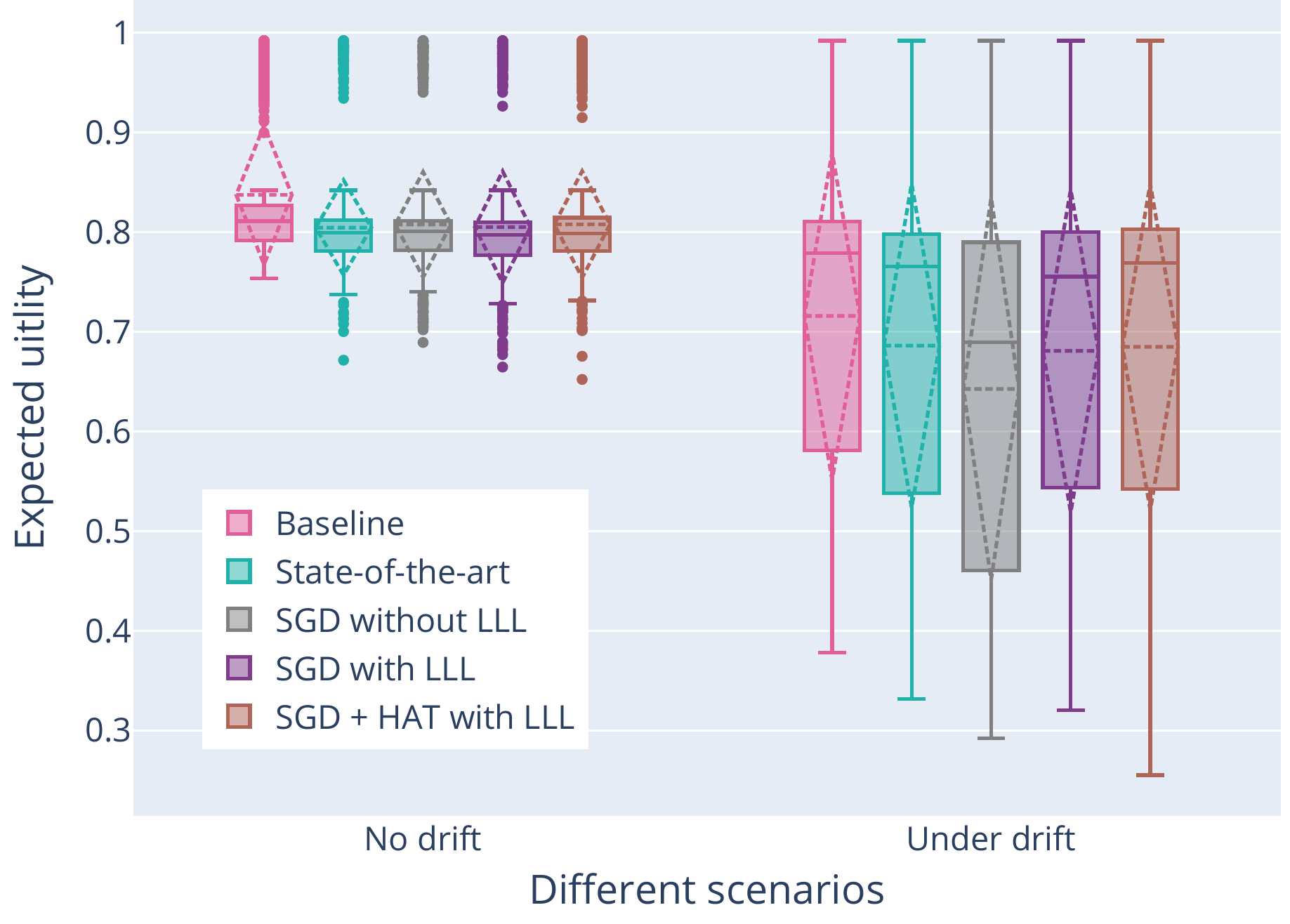}
    \caption{Utility for different methods with and without drift
    \label{fig:utility-deltaiot}\vspace{-20pt}}
\end{figure}

\subsubsection{Comparison with State-of-the-art Approach}
\begin{figure}[!h]
    \centering
    \includegraphics[scale=0.35]{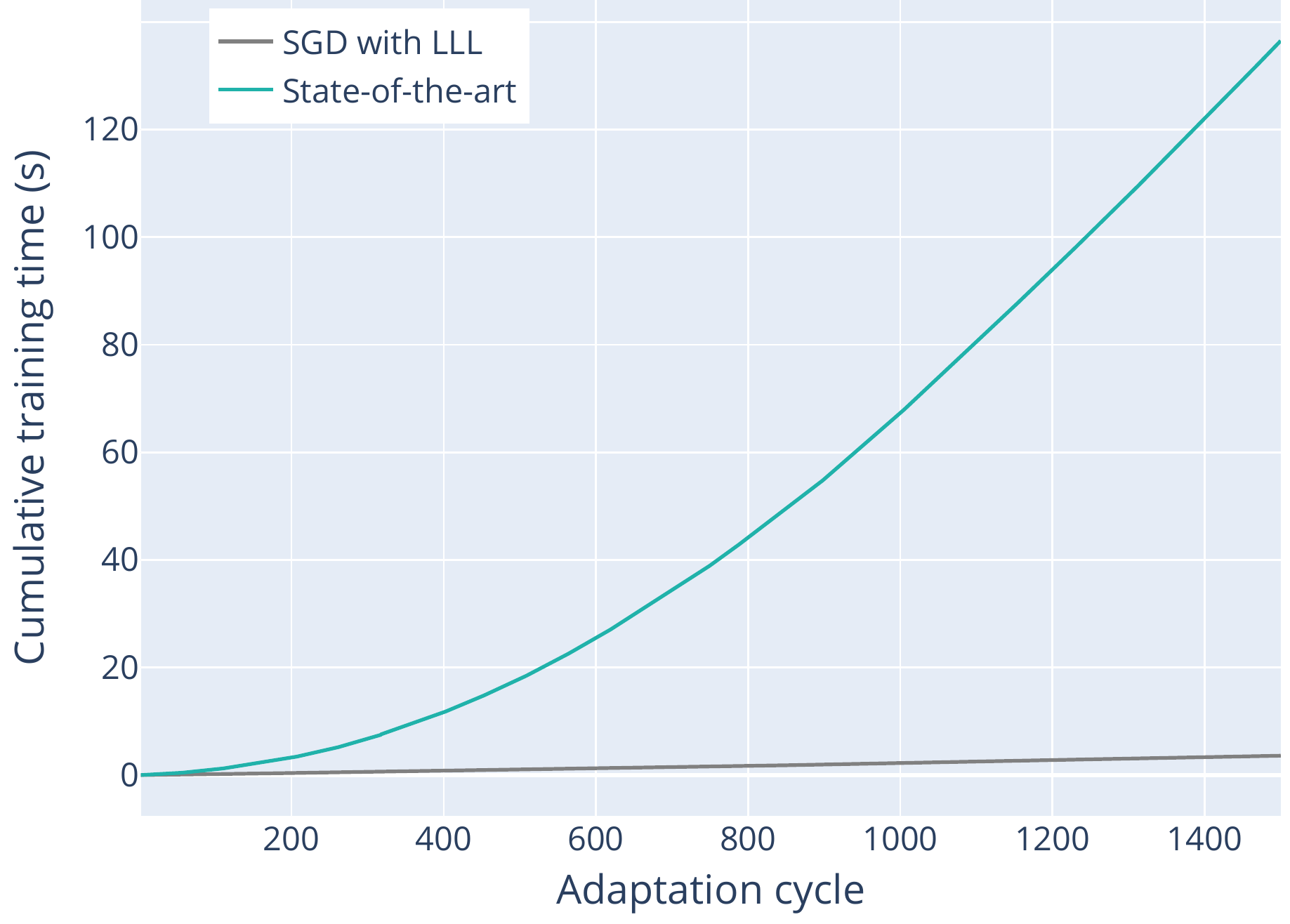}
    \caption{Cumulative training time of the state-of-the-art method (retrain the model with all existing training data) versus incremental learning under lifelong self-adaptation
    \label{fig: training time comparison}\vspace{-7pt}}
\end{figure}

When we compare the two approaches for lifelong self-adaptation (SGD with LLL and SGD + HAT with LLL) with the state-of-the-art approach~\cite{chen2019all}, we observe a very similar performance in the realization of the quality properties and overall utility of the system.  
Hence, a valid question is what the competitive advantage of lifelong self-adaptation would be compared to the state-of-the-art. The answer to this question is twofold. On the one hand, lifelong self-adaptation offers a clean conceptual approach to deal with new learning tasks (in our case tasks related to concept drift of input data of the learner). The lifelong learning loop is drift-aware and provides first-class support to identify and deal with concept drift. 
On the other hand, the state-of-the-art approach we used in this paper exploits all the previously collected knowledge to train the learner, while lifelong self-adaptation relies on incremental training where the size of the training data can be set based on the characteristics of the domain at hand. Figure~\ref{fig: training time comparison} shows the cumulative training time over the 1500 adaptation cycles for the state-of-the-art approach and SGD with LLL. We observe that the training time of the state-of-the-art approach grows much faster. Consequently, the state-of-the-art method consumes more computational resources for training the learning models. In the DeltaIoT setting, a cycle of the lifelong learning loop for SGD with LLL is only a fraction of the time window that is available to make an adaption decision at all times (i.e., 9.5 minutes). 
On the other hand, after 6500 cycles, the time required to complete an update of the learning models with the state-of-the-art approach would exceed the available time window for adaptation. Clearly, the state-of-the-art approach in its basic form is not scalable over longer periods of time. 

In answer to the second evaluation question, we conclude that lifelong self-adaptation performs similar to the  state-of-the-art approach~\cite{chen2018lifelong}. Yet, lifelong self-adaptation provides a clean conceptual approach to deal with new learning tasks that emerge from concept drift of the input data used by the learner of the managing system, and the approach scales well over time in contrast to the the state-of-the-art approach~\cite{chen2018lifelong}.

\subsubsection{Detection of new tasks}
\begin{figure}[!h]
    \centering
    \includegraphics[scale=0.35]{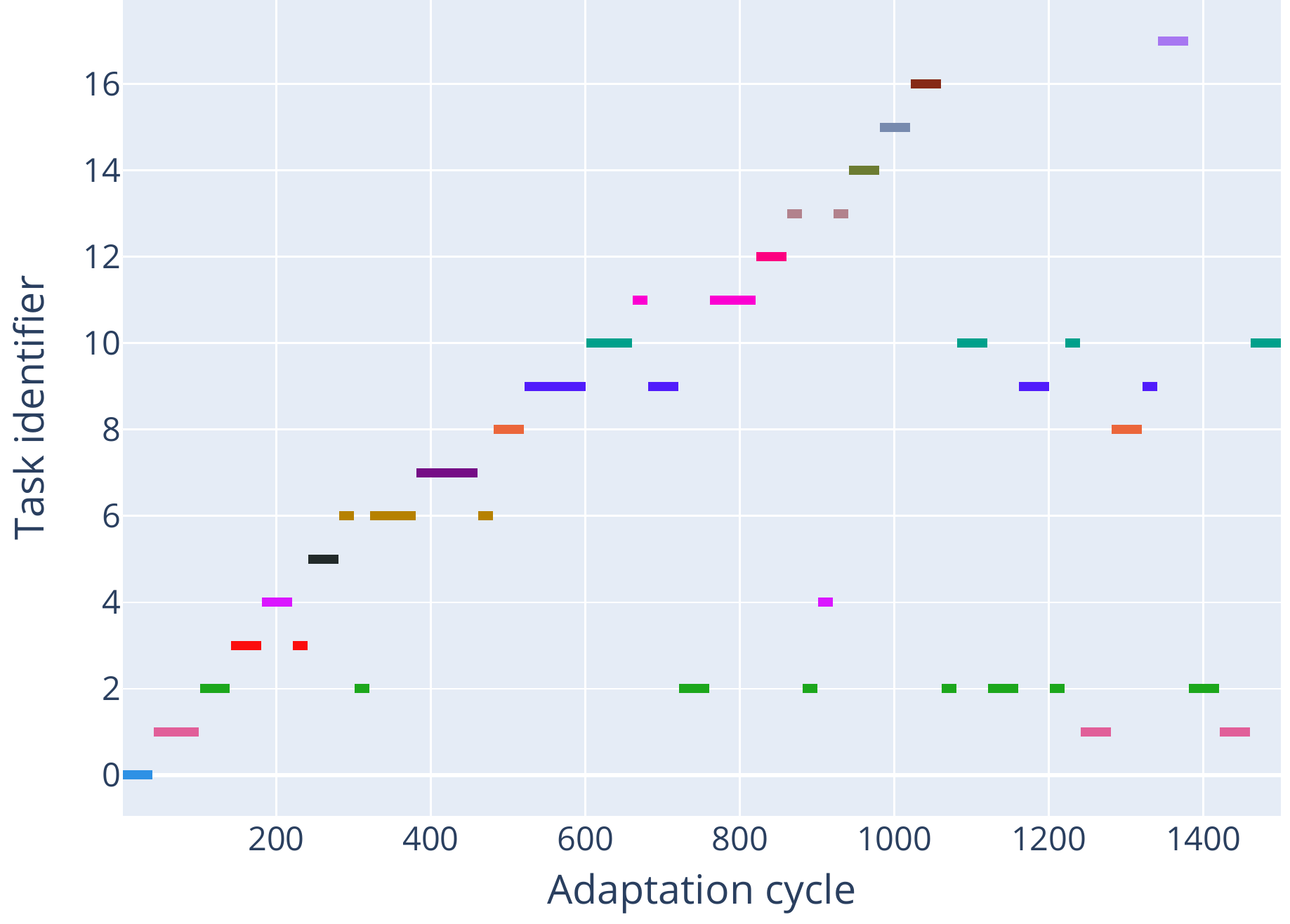}
    \caption{Tasks identified over 1500 adaptation cycles 
    \label{fig: task ids}\vspace{-10pt}}
\end{figure}

Finally, we looked at how many new tasks the lifelong learning loop detected in the evaluation scenario. Figure~\ref{fig: task ids} shows that in total \omgupdates{18} different new tasks were detected over the period of 1500 cycles. In the period from cycle 1 to cycle 600, the learning models are trained for the tasks at hand. This period includes a first period of drift (from cycle 100 to cycle 350, see Figure~\ref{fig: network interference drift}). From cycle 600 to cycle 750, when no concept drift occurs, the system evolves the learning models for some of the learning tasks based on relevant changes detected in the distributions of input data. Then from cycle 750 until cycle 1100, the system faces a new wave of drift and consequently a number of new learning tasks are identified. In the remainder of the scenario until cycle 1500 no concept drift occurs anymore. During this period, the system evolves the learning models for some of the learning tasks. Two additional tasks are detected and learned at the end of the cycle due to some gradual changes in the interference that were significant enough to define new learning tasks. 

\section{Lifelong Self-Adaptation to Deal with Incremental Co-variate Drift}
\label{sec:lifelong-self-adaptation-incremental-concept-drift} 
To explain how lifelong self-adaption deals  with sudden co-variate drift we use a case in the domain of gas delivery. 

\begin{figure*}[!h]
    \centering
    \includegraphics[scale=0.5]{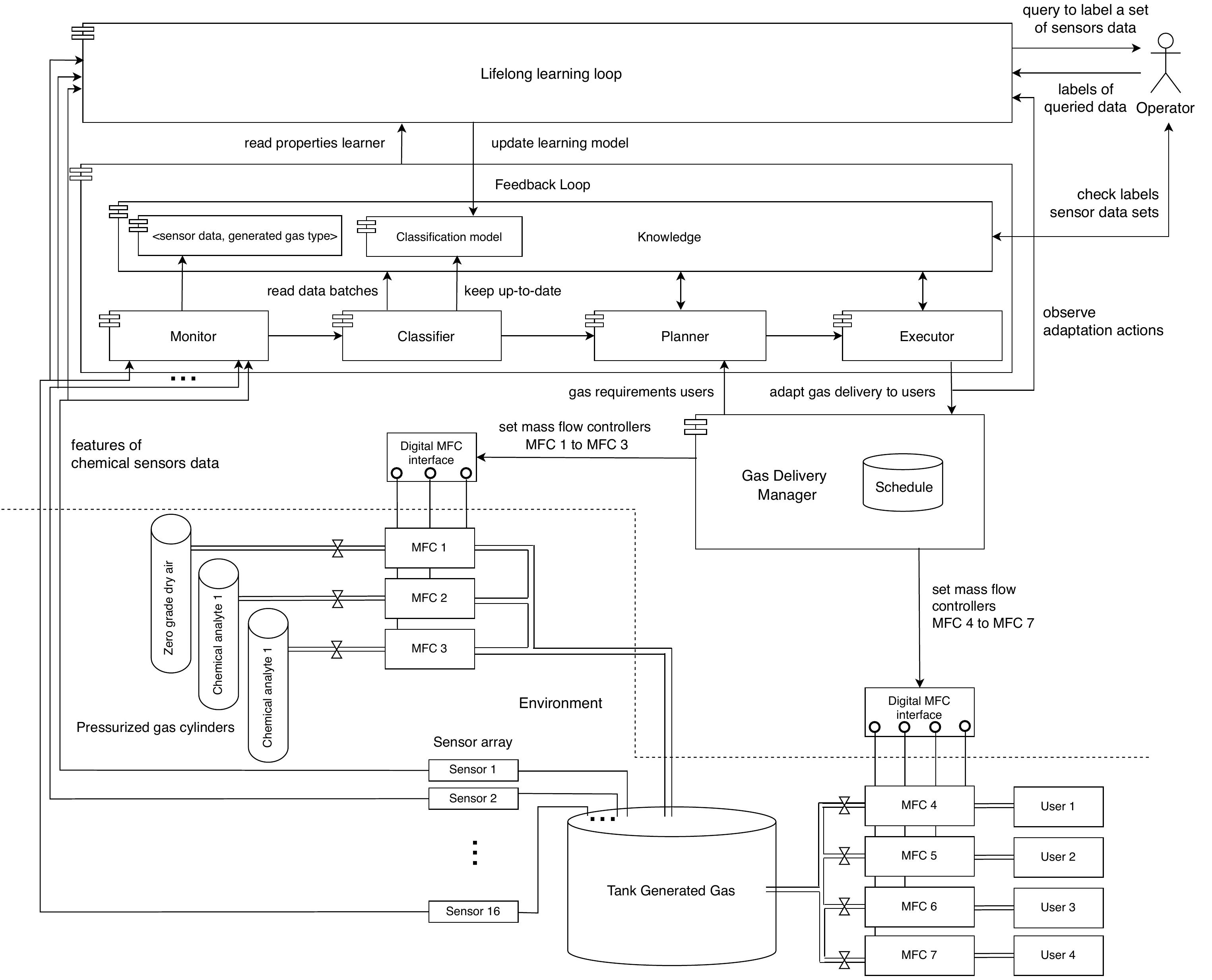}
    \caption{Gas delivery setup for the evaluation 
    \label{fig:gas-delivery-setup}\vspace{-5pt}}
\end{figure*}
Figure~\ref{fig:gas-delivery-setup} shows a schematic overview of the setting that is inspired by~\cite{vergara2012chemical}. 
The schedule of the system determines the type of gas that needs to be produced and the user requirements for a given period of time. Different types of gas can be generated by routing different portions of pressurized chemicals to the tanker. The gas delivery manager manages the mass flow controllers MFC1-3 to generate gas according to the schedule. Yet, there is uncertainty about the type of gas that is produced, which needs to be resolved before gas is routed to the users. To that end, the feedback loop collects sensor data from an array of 16 sensors at the gas tank and stores the data in the knowledge repository. The classifier, a multi-class support vector machine (SVC)\footnote{SVC combines several binary support vector machines based on a given strategy~\cite{angulo2003k}. Example strategies consider decision boundaries between training examples of every two classes (``one vs. one'' strategy) or training examples of each class against all others (``one vs. all'' strategy).}, 
labels the sensor data with a gas type, yielding tuples $\langle$\textit{sensor data, generated gas type}$\rangle$. Periodically, the operator checks the labels and corrects them if needed (batches of 100). The classifier uses this input to improve its performance over time. Based on the gas type of the current data tuple and the gas requirements of the users, the planner determines the delivery of gas to the users. The executor then enacts these settings to the gas delivery manager that sets the valves of the mass flow controllers MFC4 to MFC7 routing the gas to the right users. 

\subsection{Problem}

\begin{figure}[!h]
    \centering
    \includegraphics[scale=0.4]{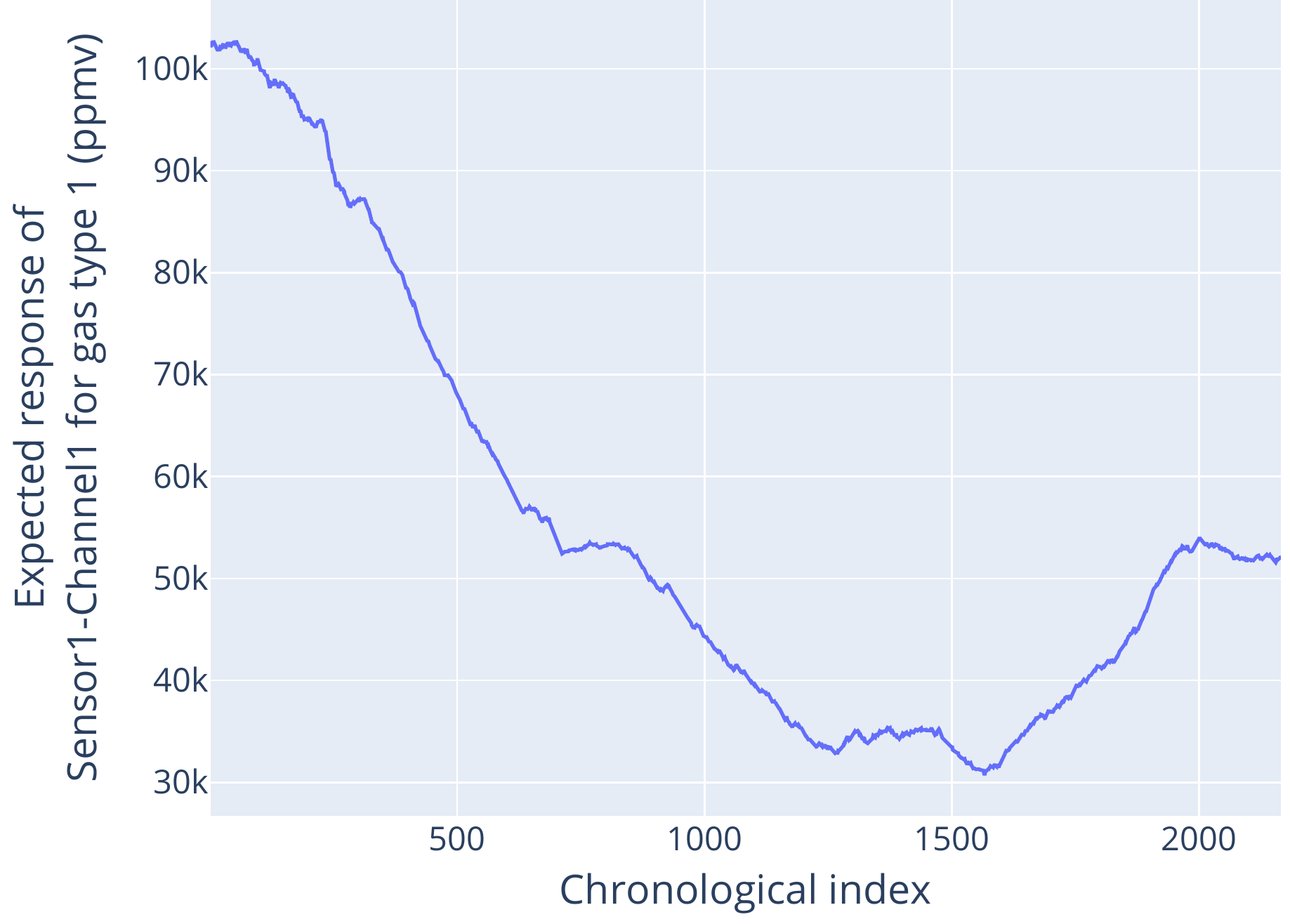}
    \caption{Moving average (with 400 cycles window size) of data from one sensor showing incremental co-variate drift 
    \label{fig: GDS drift scenario}\vspace{-10pt}}
\end{figure}

A problem with this setting is the potential degradation of the surface of the sensors due to ageing or contamination, causing sensor drift, a challenging problem in chemical sensing~\cite{vergara2012chemical}. Figure~\ref{fig: GDS drift scenario} shows the moving average (400 cycles window) of the data obtained from one of the sensors, illustrating the drift of data over time. This type of drift, characterized as incremental co-variate drift, may affect the performance of the learner over time. To deal with this problem, we add a lifelong learning loop on top of the feedback loop, see Figure~\ref{fig:gas-delivery-setup}.  

\subsection{Lifelong Learning Loop Instance}

We empirically determined that lifelong learning is triggered best every 20 adaptation cycles of the managing system. 

\subsubsection{Knowledge Manager} Collects 20 new knowledge triplets per cycle of the lifelong learning loop. Here, input of the triplets are features of the sensor array data. State is meta-data about learner, including the configuration of the learner and sets of pairs of sensory data with labels checked per 100 cycles by the operator. 
Output are the adaptation actions applied based on the classification label of the sensory data. In this instance, the knowledge manager does not maintain any meta-knowledge. 

\subsubsection{Task Manager} Uses the same algorithm of the task manager in the first validation case. Yet, the similarity threshold is relaxed to 0.05 (in contrast to 0.025 in the first case). 

\subsubsection{Knowledge-Based Learner} Evolves the learning models using an algorithm similar to Algorithm~\ref{algo:knowledge based learner}. 
Based on the data received from the task-based knowledge miner, the knowledge-based learner determines the best learning model (optimizing the loss value for the learner with a Bayesian optimizer). Then, it trains the best model and updates the classification model in the knowledge repository of the managing system.

\subsubsection{Task-Based Knowledge Miner} Fetches the sets of knowledge triplets from the knowledge manager for the tasks queried by the knowledge-based learner. Due to uncertainties in the gas composition, the labels of the input in the state of these triplets may not necessarily be correct. To that end, for new detected tasks, the task-based knowledge miner selects the top five\footnote{This number was  experimentally determined.} inputs (out of 20 newly observed triplets) with the highest uncertainty.\footnote{The uncertainty is proportional to the inverse of the standard deviation of distances between the input and decision boundaries of SVC~\cite{guo2015active}.} 
Then, task-based knowledge miner asks the operator to check these labels 
and once it receives the corrections, sends them to the knowledge manager to use them for improving the performance of the learner. 
If the detected task is not new, the task-based knowledge miner fetches a sample of knowledge triplets for the detected task.\footnote{The size of this sample can be determined based on the capacity of the learning model~\cite{gheibiSASCLT2021}. We used up to 1000 samples for the evaluation case.} The state of the triplets contains correctly labeled data that can be used by the knowledge-based learner for updating the learning model.

\vspace{-5pt}
\subsection{Evaluation Results}

To validate lifelong self-adaptation for incremental co-variate drift, we implemented a simplified simulation of the gas delivery setup shown in Figure~\ref{fig:gas-delivery-setup}. Concretely, we implemented the feedback loop, the lifelong learning loop, and the part that provides gas to users. The remainder of the system and the environment are represented as an extensive set of data~\cite{gasdataset}. 
This data set was generated over three years in a controlled setting; it contains the sensor data and the correct labels for the operator (ground truth). 
The classifier of the feedback loop was realized as an ensemble of classifiers based on support vector machines. The weighted combination of these classifiers were trained at different points in time to solve a gas discrimination problem with high accuracy. 
We used the same computer setup as for the first case for the evaluation. 

\begin{figure}[!h]
    \centering
    \includegraphics[scale=0.43]{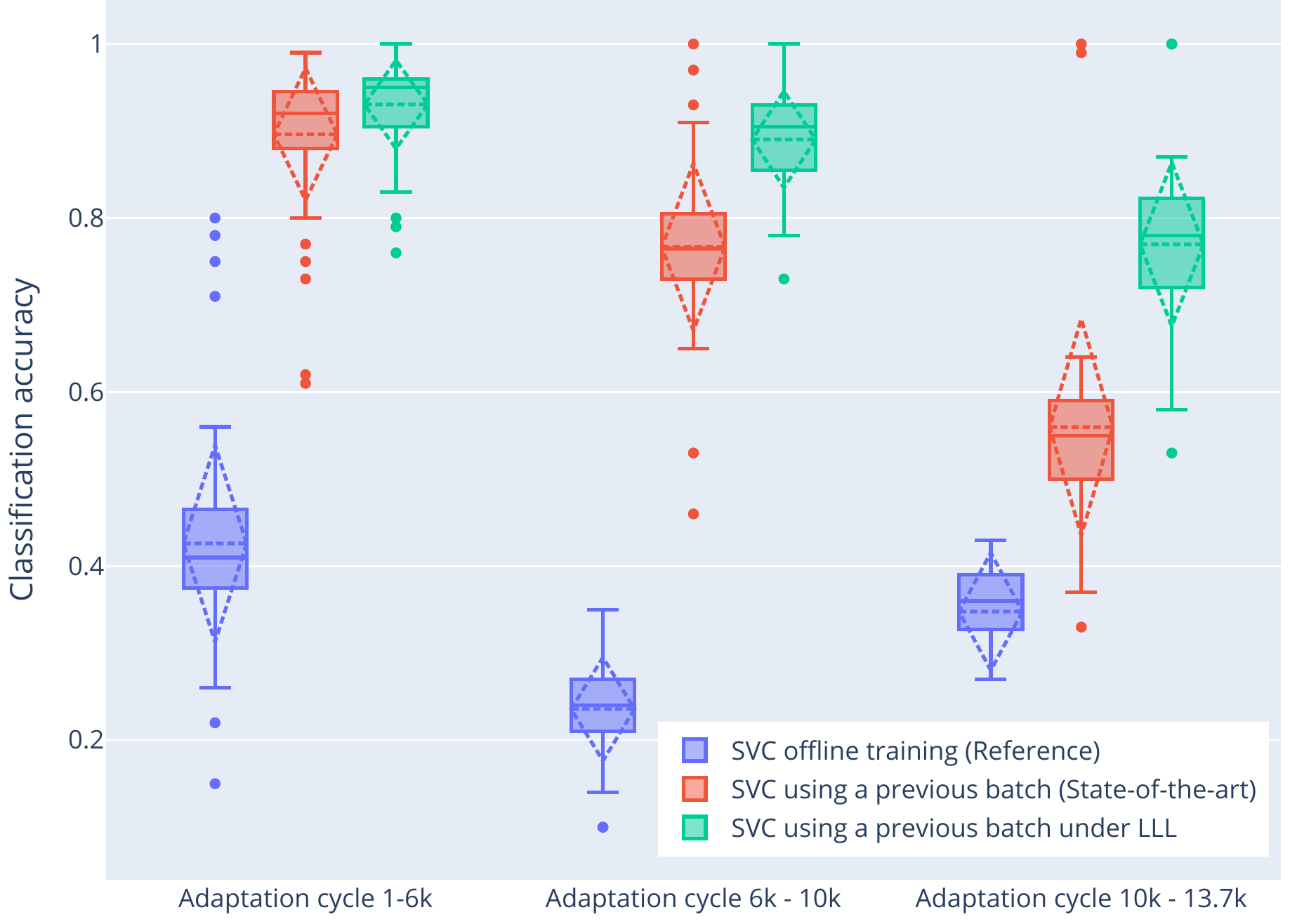}
    \caption{Classification accuracy of the different approaches    \label{fig: classfication accuracy in three time ranges}\vspace{-20pt}}
\end{figure}

We compared three learners: (1) a SVC trained offline (Reference), (2) a SVC with online training that uses the data of the previous batch of monitored data (State-of-the-art~\cite{VERGARA2012320}), and (3) the SVC of (2) with a lifelong learning loop. We measured the classification accuracy using the ground truth that is provided with the data set~\cite{gasdataset}. Figure~\ref{fig: classfication accuracy in three time ranges} shows the results over 13.7k cycles, grouped in three zones. The results show low values for accuracy with the reference approach in all zones. While the state-of-the-art approach scores similar to the approach with a lifelong loop in the first zone, its accuracy gradually decreases afterwards. The mean value of the accuracy over the complete run was $0.35$ for the reference approach, $0.77$ for the state-of-the-art, and $0.88$ for the lifelong self-adaptation approach. The results show  that lifelong self-adaptation can effectively deal with incremental concept drift. 

\section{Threats to Validity}\label{sec:validity}

The evaluation of lifelong self-adaption is subject to a number of validity threats.
We evaluated the approach only for two types of concept drift, so we cannot generalize the findings for other types of problems that require dealing with new learning tasks (external validity). Additional research is required to study the usefulness of the architecture for other challenges with new learning tasks in self-adaptive \omgupdates{systems}. Additionally, we validated the architecture for each type of concept drift with a single application. Evaluation in different domains is required to increase the validity of the results for the types of concept drift considered in this paper. 
We have evaluated the instances of the architecture for particular settings. The type and number of new learning tasks these settings generate may have an effect on the difficulty of the problems (internal validity). We mitigated this threat by instantiating the architecture for two different domains. However, additional evaluation in other domains is required to increase the validity of the results for the types of concept drift studied. 
For practical reasons, we used simulation for the evaluation with data that contains uncertainty. The results may be different if the study would be repeated
(reliability). We minimized this threat by considering data extracted from real systems, and we evaluated the cases over long periods of time. We also provide a replication package for the study~\cite{project-website}. 

\section{Related Work}\label{sec:related-work}

We look at a selection of work at the crossing of machine learning and self-adaptation, focusing on concept drift and performance of machine learning in self-adaptive systems. 
\vspace{5pt}\\
\noindent
\textbf{Dealing with Concept Drift in Self-Adaptation}. 
T. Chen~\cite{chen2019} studied two methods to deal with concept drift in  self-adaptation with learning: retraining a new model in each cycle using all available data, and retraining the existing model using the newly arrival data sample. In contrast to general beliefs about the choices between the two, the author examines both modeling methods for distinct domains of adaptive software and identifies evidence-based factors that can be used to make well-informed decisions. 
Bierzynski et al.~\cite{Bierzyn2019ski} present the architecture of a self-learning lighting system that equips a MAPE-K loop with a learner that learns activities and user preferences. Another feedback loop on top of the learner determines when the predictions of a learning model start to drift and then adapts the learning model accordingly. 
A concrete instance based on micro-services is proposed and compared with other architectural styles. 
Vieira et al.~\cite{vieira2021} propose Driftage, a multi-agent framework for concept drift detection that uses a MAPE-K loop. The monitor and analyzer agents capture and predict concept drifts on data, and the planner and executor agents determine whether the detected concept drift should be alerted. The approach is applied to health monitoring of muscle cells. 
Casimiro et al.~\cite{Casimiro0GMKK21} propose a framework \omgupdates{for self-adaptive} systems with machine-learned components. The authors outline
a set of adaptation tactics to deal with misbehavior of \omgupdates{the machine-learned components}, the required changes to the MAPE-K loop for dealing with them, and the challenges associated with developing this framework.

\begin{sloppypar}
In contrast to these approaches, we provide a domain-independent architecture to deal with new emerging tasks in learning modules used by self-adaptive systems. The work of~\cite{chen2019} and~\cite{Casimiro0GMKK21} offer valuable solutions that can be used when instantiating the architecture for lifelong self-adaptation. 
\end{sloppypar}

\vspace{5pt}
\noindent
\textbf{Improving the Performance of Machine Learning for Self-Adaptation}. 
Jamshidi et al.~\cite{jamshidi2018} propose L2S (Learning to Sample), an efficient approach for transferring knowledge across execution environments to simplify the configuration, e.g., hardware, software release. The approach progressively concentrates on interesting regions of the configuration space. 
Chen and Bahsoon~\cite{Chen2017} present a modeling approach for creating quality prediction models that use data from the environment and control settings as inputs. The approach leverages adaptive multi-learners, selecting the best model for prediction on the fly, and is evaluated in a cloud environment. Chen et al.~\cite{chen19} apply a similar approach to deal with the problem if resource allocation for cloud-based software services.
Chen et al.~\cite{chen2010experience} propose the idea of ``experience transfer’’ from one system to another similar system, and demonstrate this for system configuration tuning, where  experiences are dependencies between configuration parameters. 

These related approaches target the efficiency of machine learning methods in the context of self-adaptation. Our work complements these approaches focusing on enhancing learning to handle new learning tasks as required for concept drift. 

\section{Conclusions and Future Work}\label{sec:conclusions} 
Self-adaptive systems are characterized by their ability to deal with uncertainty. Machine learning techniques on the other hand, which are increasingly used in self-adaptive systems, face challenges with uncertainties that were not considered as learning tasks during design. To deal with new learning tasks, we presented lifelong self-adaptation, a novel approach to self-adaptation that enhances self-adaptive systems that use machine learning techniques with a lifelong learning layer. We presented an architecture for lifelong self-adaptation and applied it to deal with two types of concept drift in two domains. The evaluation results show that lifelong self-adaptation effectively resolves the problems of concept drift. 

\begin{sloppypar}
In future work, we plan to study and apply lifelong self-adaptation for other types of concept drift, in particular class drift and novel class appearance. Then, we plan to investigate how lifelong self-adaptation can be used to deal with new and changing adaptation goals, which is a particularly challenging example of uncertainties for self-adaptive systems. 
\end{sloppypar}

\newpage
\bibliography{main}
\bibliographystyle{acm}
\end{document}